\documentclass[twocolumn,twocolappendix,fleqn]{aastex631}

\usepackage{tikz}
\usepackage{CJK}
\usepackage{multirow}

\usepackage{amsmath,amssymb,amstext}
\usepackage{graphicx}
\usepackage{enumitem}
\usepackage{natbib}
\usepackage{tablefootnote}

\newcommand{\br}{\ensuremath{\mathrm{BP-RP}}}
\newcommand\MH{\ensuremath{\mathrm{[M/H]}}}
\newcommand\nsample{533,584}

\DeclareMathOperator*{\argmax}{argmax}

\definecolor{mygreen}{HTML}{006d2c}
\definecolor{myblue}{HTML}{0868ac}
\definecolor{mybluedark}{HTML}{74a9cf}
\definecolor{mybleuclair}{HTML}{f7fbff}
\definecolor{mypurple}{HTML}{6a51a3}
\definecolor{mybrown}{HTML}{67001f}
\definecolor{mypink}{HTML}{980043}
\definecolor{mypinkclair}{HTML}{f7f4f9}
\definecolor{myorange}{HTML}{fe9929}
\definecolor{myorangeclair}{HTML}{fff7ec}
\definecolor{mypurpleclair}{HTML}{fcfbfd}
\definecolor{mybluea}{HTML}{084081}

\shorttitle{Mono-age mono-abundance maps of the LMC}
\shortauthors{Frankel et al.}

\begin{document}

\author{Neige Frankel}
\affiliation{Canadian Institute for Theoretical Astrophysics, University of Toronto, 60 St. George Street, Toronto, ON M5S 3H8, Canada}
\affiliation{David A. Dunlap Department of Astronomy and Astrophysics, University of Toronto, 50 St. George Street, Toronto, ON M5S 3H4, Canada}
\author{Rene Andrae}
\affiliation{Max Planck Institute for Astronomy, K\"onigstuhl 17, D-69117 Heidelberg, Germany}
\author{Hans-Walter Rix}
\affiliation{Max Planck Institute for Astronomy, K\"onigstuhl 17, D-69117 Heidelberg, Germany}
\author{Joshua Povick}
\affiliation{Max Planck Institute for Astronomy, K\"onigstuhl 17, D-69117 Heidelberg, Germany}
\author{Vedant Chandra}
\affiliation{Center for Astrophysics $\mid$ Harvard \& Smithsonian, 60 Garden St, Cambridge, MA 02138, USA}

%
%
%
%
%
%
\title{What Does the Large Magellanic Cloud Look Like? It Depends on [M/H] and Age}

%
%
%
%
%
%

\begin{abstract}
We offer a new way to look at the Large Magellanic Cloud through stellar mono-abundance and mono-age-mono-abundance maps. These maps are based on $\gtrsim 500\,000$ member stars with photo-spectroscopic [M/H] and age estimates from Gaia DR3 data, and they are the first area-complete, metallicity- and age-differentiated stellar maps of any disk galaxy. Azimuthally averaged, these maps reveal a surprisingly simple picture of the Milky Way's largest satellite galaxy. For any [M/H] below -0.1 dex, the LMC's radial profile is well described by a simple exponential, but with a scale length that steadily shrinks towards higher metallicities, from nearly 2.3~kpc at [M/H]$=-1.8$ to only 0.75~kpc at [M/H]$=-0.25$. The prominence of the bar decreases dramatically with [M/H], making it barely discernible at [M/H]$\lesssim -1.5$. Yet, even for metal-rich populations, the bar has little impact on the azimuthally averaged profile of the mono-abundance components. Including ages, we find that the scale length is a greater function of age than of metallicity, with younger populations far more centrally concentrated. At old ages, the scale length decreases with increasing metallicity; at young ages, the scale-length is independent of metallicity. These findings provide quantitative support for a scenario where the LMC built its stellar structure effectively outside in.
\end{abstract}


%
%
%
%
%
%
\section{Introduction}

The Large Magellanic Cloud (hereafter LMC) is the largest satellite galaxy of the Milky Way \citep[e.g.][]{2009MNRAS.392L..21S}, presumably on its first infall towards the Milky Way \citep{2007ApJ...668..949B, besla_2012}. The LMC has a stellar mass of $2\times 10^{9}$M$_\odot$ \citep{vanderMarel2002} and total mass of $2.5\times 10^{11}$M$_\odot$ \citep{Penarrubia2016}. It has a disk, spiral arms and a prominent central bar \citep[e.g.][]{2013A&A...552A.144S}. Its distance from us is about $50\,\mathrm{kpc}$ \citep{2019Natur.567..200P, van_der_marel_cioni_2001, de_grijs_2014}, and it has experienced close tidal interactions with the Small Magellanic Cloud \citep[hereafter SMC,][]{2016ApJ...825...20B} with which it shares a very similar star-formation history \citep{Harris_2009}.

Although not as massive as M31 or M33, the LMC is 15 times closer than these galaxies \citep{van_der_marel_2012b}, making it a unique target to study an entire sizeable galaxy star-by-star, mapping and understanding its structure and kinematics as a function of the stars' ages and metallicities.

Until recently, only small samples of spectroscopic measurements were available for the LMC \citep[e.g.][]{2005AJ....129.1465C,2008A&A...480..379P,2012ApJ...761...33L,2013A&A...560A..44V,2017AJ....153..261S} with, at the time of this analysis, the largest sample comprising $\sim$5000 RGB stars from SDSS-IV (APOGEE) in both the LMC and SMC \citep{2020ApJ...895...88N}. 
These have been used by \citet{2021ApJS..252...23S} and \citet{povick_2023b_abundance_gradients} to map the metallicity gradients in the LMC.
An extended version of the SDSS-IV sample is currently being analyzed for more in-depth studies of the age-metallicity structure \citep{povick_2023b_abundance_gradients}. 
The ongoing LMC program of SDSS-V \citep{Kollmeier2017} and the planned 1001MC program of the 
4MOST survey \citep{2019Msngr.175...54C} will provide ground-based spectra and 
abundances for hundreds of thousands of stars in the LMC.
Furthermore, in the next Gaia data release (DR4), RVS spectra should reach fainter and may also cover hundreds of thousands of LMC member stars.

As an alternative to spectroscopy, photometry can be used to constrain metallicities
for extensive samples of stars by matching their colors and luminosities with isochrones \citep[e.g.][]{2017AJ....154..199N}, especially if near-infrared is used \citep[e.g.][]{Piatti_2013,10.1093/mnras/stv2414,2021MNRAS.507.4752C,2021ApJ...909..150G}. However, with purely photometric approaches, there will inevitably be a substantive covariance between ages and metallicities, at least for giant stars.

However, already now there is a vast sample of stars ($>$500,000) available that covers the entire face of the LMC with spectroscopic stellar parameters and metallicities: Gaia's low-resolution XP spectra \citep{2021A&A...652A..86C,2022arXiv220606143D,GaiaDR3}. Despite their modest spectral resolution (between 20 and 50) these XP provide robust and precise stellar metallicity estimates for millions of cool giant stars across the sky \citep{2022ApJ...941...45R,Andrae_rix_chandra_2023,2023arXiv230303420Z,2023arXiv230317676Y}. As of now, stellar XP spectra have been published only for sources brighter than $G<17.65$, which leads to a luminosity limit of $M_G\lesssim -0.85$) at the LMC's distance. Once we know the metallicites and effective temperatures, $T_\mathrm{eff}$ of luminous cool stars from spectroscopy, their ages can be constrained well if precise distances and photometry are also available, as is the case for the LMC \citep[e.g.][]{povick_2023a_ages}.

Studying the structure of different mono-abundance and mono-age stellar populations has proven a powerful tool for understanding the formation and evolution history of the Milky Way \citep[e.g.][]{bovy_etal_2012,rix_bovy_2013,frankel_etal_2019,frankel_2020}. Here, we present a first study of the stellar structure of the LMC, looking only at stars of only a given metallicity or age. These stellar labels serve as (largely) immutable birth tags that separate different evolutionary sub-populations. This approach preserves more information than considering only the mean metallicity at each radius, i.e. the metallicity gradient, as done with a similar sample in complementary work by \cite{massana_2024}.

We focus on the the azimuthally averaged mono-abundance and mono-age structure of the LMC, although the LMC is manifestly not axisymmetric at present. It has a central bar that is off-center and a single arm spiral \citep{devaucouleur_1955, devaucouleur_1972, zaritsky_2004}, it is lopsided and its geometric and kinematic centers do not coincide \citep[e.g.,][]{van_der_marel_2001, cole_2005}, it is warped \citep{choi_2018}. Many of these asymmetries may be attributable to the recent interaction with the SMC.

To do this in practice, we draw on the \MH\ estimates that \citet{Andrae_rix_chandra_2023} derived from XP spectra and CatWISE photometry \citep{2021ApJS..253....8M}. And, we draw on the stellar ages, $\tau_*$, for these stars, derived by Povick et al (\emph{in prep.}). In Sect.~\ref{sec:data-and-selection}, we recap these \MH\ and $\tau_*$ estimates, define our selection of LMC members and validate the \MH\ estimates against spectroscopic values specifically from the LMC. In Sect.~\ref{sec:global-MH-inspection} we inspect the global properties of the \MH\ distribution in the LMC before we model the metallicity-dependent structure in more detail in Sect.~\ref{sec:spatial-structure-modelling}. In Sect.~\ref{sec:spatial-structure-modelling-age-mh}, we include the age dimension and model the age-metallicity-dependent structure of the LMC. We discuss our findings in Sect.~\ref{section:discussion}.

%
%
%
%
%
%
\section{Data and Selection}
\label{sec:data-and-selection}

\subsection{\MH\ estimates for the LMC}

\citet{Andrae_rix_chandra_2023} have published an extensive catalog which also covers the LMC. In slight variation of this work, we devise a new set of \MH\ estimates with two adaptations: uncertainty estimates and accounting in the analysis for the known LMC distance.

The uncertainty estimation is done by drawing Monte-Carlo samples from the Gaussian measurement errors of the XP coefficients and the Gaia and CatWISE photometry. For each source, 100 Monte-Carlo samples are drawn and processed through the exact same XGBoost models trained in \citet{Andrae_rix_chandra_2023}. This results in 100 Monte-Carlo estimates of \MH\ (and $T_\mathrm{eff}$ and $\log g$). These are then summarised by a mean value and a standard deviation, thus providing an uncertainty estimate.

While \citet{Andrae_rix_chandra_2023} designed the XGBoost input features to depend linearly on parallax in order to minimise the impact of parallax noise, the parallaxes of LMC member stars have very low signal-to-noise ratios in Gaia~DR3. This will inevitably confuse the XGBoost estimation. In order to overcome this limitation, we adopt the known LMC distance taking into account the orientation and inclination of the LMC disk as defined in \citet{2023arXiv230606348P}. We then replace the parallax by the inverse distance in the input features and again use the same XGBoost models as trained in \citet{Andrae_rix_chandra_2023}. We find that replacing the noisy parallax by the inverse LMC distance reduces the \MH\ uncertainties by $\sim$15\%. The remaining \MH\ uncertainties originate from the Gaia and CatWISE photometry as well as the XP spectra.
We provide an extensive validation of these \MH$_{XP}$\ estimates against extant (mostly high-resolution) spectroscopy in Appendix~\ref{appendix:XPvalidation}.

\subsubsection{\MH$_{XP}$\ Completeness}
\begin{figure}
\includegraphics[width=0.47\textwidth]{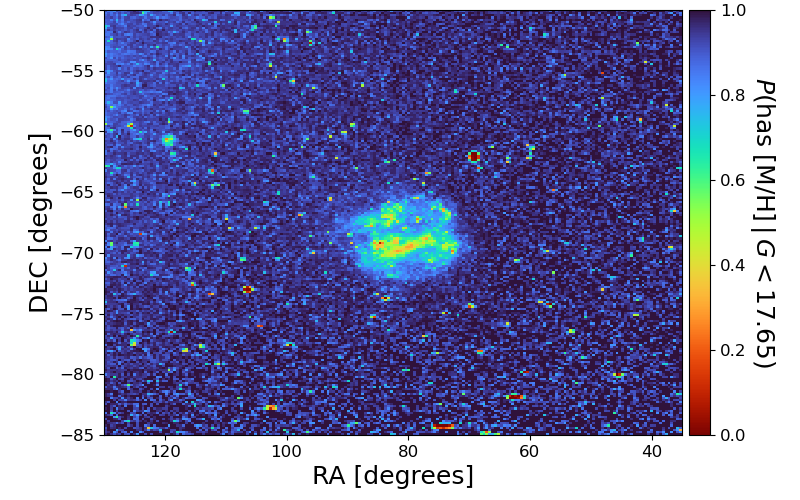}
\caption{Probability to have \MH\ estimates from XGBoost for a star with apparent magnitude $G<17.65$ in the LMC footprint. This completeness map is derived in Appendix~\ref{appendix:completeness} and following Eq.~\ref{eq:completeness-factorised}, it is the product of the four separate maps shown in Fig.~\ref{fig:completeness-step-by-step}.}
\label{fig:LMC-completeness}
\end{figure}

Understanding the selection function of LMC stars that end up having \MH$_{XP}$\ values is critical for our subsequent structure analysis. To have a valid \MH$_{XP}$\  mainly requires a star a) to be brighter than G$=17.65$ (or else no XP information is published), b) not to be affected by source crowding in the Gaia data (as the XP spectra are slitless spectra that overlap more easily among sources than the undispersed images), and c) have valid CATWISE photometry (where crowding is more severe due to the large PSF of the WISE satellite). 

We approach this problem by presuming that the selection function, $p( \mathrm{has\,} \MH\, | G>17.65 \,\mathrm{mag})$, can be described as the product of four terms: whether stars $G<17.65$ mag are in the Gaia catalog at all; if so, whether they have published XP spectra; if so, whether they have CATWISE W$_{1/2}$ values; and finally, if so, whether they have an \MH$_{XP}$\ entry. This must be mapped as a function of sky position, as the completeness will -- in detail -- depend both on the source crowding and the Gaia scanning law. This is laid out in Appendix~\ref{appendix:completeness} with its Figure~\ref{fig:completeness-step-by-step}. 

The resulting overall completeness map, $p\bigl (\ \mathrm{has}\, \MH\ |\  \alpha,\delta,\  G<17.65\bigr )$, for stars brighter than $G<17.65$ in the LMC footprint is shown in Fig.~\ref{fig:LMC-completeness}. The Figure shows that the selection function is near unity in most regions across the LMC patch, but drops considerably in the central bar region, which is the densest in sources.

\subsubsection{Subsample selection for robust and precise stellar parameters}
\begin{figure}
\includegraphics[width=0.47\textwidth]{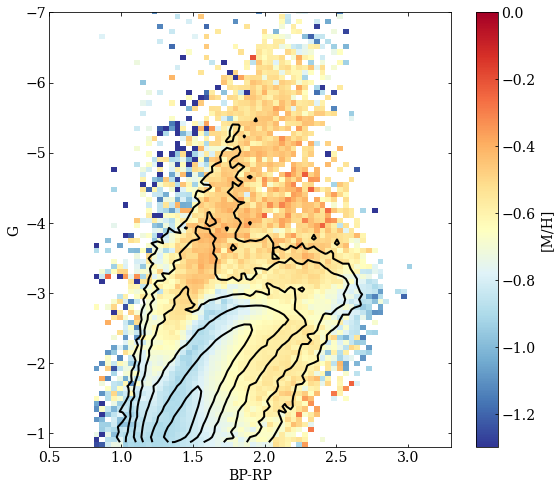}
\caption{Colour-vs-apparent-magnitude diagram of LMC member stars colour-coded by mean [M/H]. Contours indicate number density, spaced by factors of 2.}
\label{fig:LMC-MH-in-CMD}
\end{figure}

For two reasons the robustness and precision of these \MH$_{XP}$\ values depend not only on the signal to noise of the XP spectra, but also on the stars' position in the T$_\mathrm{eff}$ (or color) -- luminosity plane. 
First, there is the astrophysical reason that metallicity estimates from low-resolution spectra rely on deep and broad absorption features such as the CaII (\@ 4000\AA ) or 
the Mg\emph{b} feature at 5100\AA , which are only prominent in cool stars, say $T_\mathrm{eff}<5000$~K. Second, the \MH$_{XP}$\  estimates are data-driven and rely on dense and accurate coverage 
in the training set. Both at hot temperatures, and at very cool temperatures $T_\mathrm{eff}\lesssim 3500$~K our training set is not optimal in this respect.
Therefore, we proceed with only a subsample in the following analysis, with quality cuts designed to assure robust and precise \MH$_{XP}$s. Guided by the analysis in \citep{Rix2022}, we adopt the following colour-magnitude (Eq.~\ref{eq:good-MH-cuts-1}) and quality (Eq.~\ref{eq:good-MH-cuts-2}) cuts that afford a wide range of metallicities and a wide range of ages ($7.5<\mathrm{log}\tau_*<10.1$):
\begin{equation}\label{eq:good-MH-cuts-1}
    \begin{split}
        &G_\textrm{BP}-G_\textrm{RP} > 0.8,\\
        &1.5 <  G - W_1 < 4.2~,
    \end{split}
\end{equation}
and
\begin{equation}\label{eq:good-MH-cuts-2}
    \begin{split}
    &3450\mathrm{\,K} <  T_\mathrm{eff} <4800\, \mathrm{K},\\
    &\mathrm{RUWE} < 1.5.
    \end{split}
\end{equation}

The resulting sample is shown in Figure~\ref{fig:LMC-MH-in-CMD}. In absolute magnitude that sample is simply limited by the Gaia-based cut $G<17.65$ required for the DR3 publication of publication XP spectra. This Figure illustrates that at 50kpc distance, the red clump is already too faint to have public XP spectra for the LMC; we are only left with the upper giant branch. However, these quality cuts include stars also the red side of the ``blue loop'' that young massive stars undergo in their evolution: this provides metallicities also for very young stars ($\log_{10}(\tau_*/\mathrm{yr})<8$).

For the full analysis, we presume that this sub-sample selection, $S_{sub}(T_\mathrm{eff},\mathrm{colors})$ can simply be multiplied with the samples basic selection function $p\bigl (\mathrm{has}\, \MH\ |\  \alpha,\delta,\  G<17.65\bigr )$.

\begin{figure*}
\includegraphics[width=\textwidth]{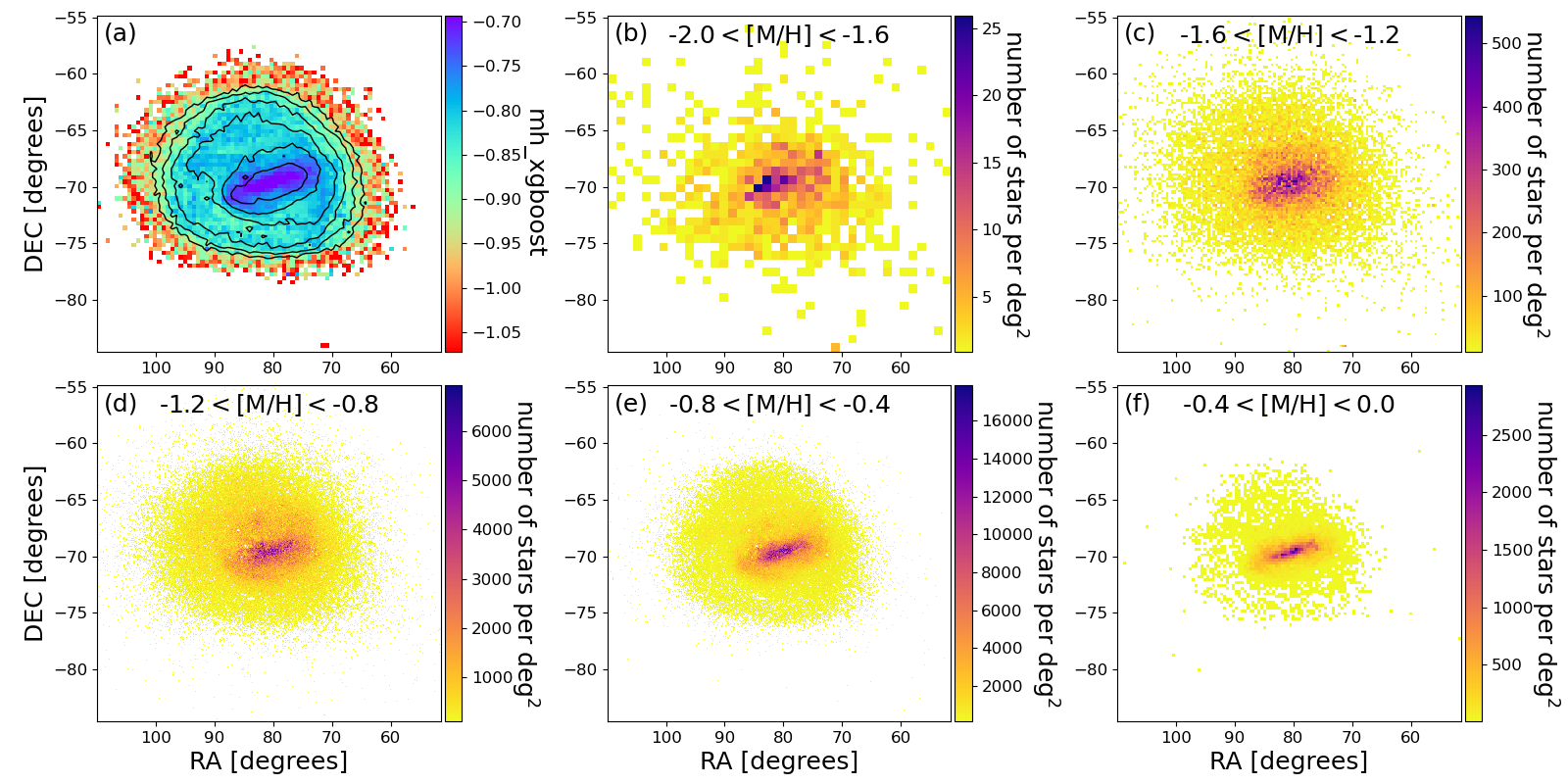}
\caption{Metallicity maps of the LMC. Panel (a) shows the global map of the mean metallicity (color coded), with total number density contours that are spaced by factors of two. Panels (b)-(f) show the completeness-corrected maps of various mono-abundance populations with \MH\ ranging from $-2$ to 0. All panels mask noise-dominated pixels that contain fewer than 5 stars.}
\label{LMC-mono-abundance-maps}
\end{figure*}

\subsection{Kinematic and Spatial Selection of LMC members}

We now want to use additional information to assign to each star in the LMC vicinity a membership probability. This is detailed in 
Frankel+24 (\emph{in prep.}) and summarized briefly here. We construct empirical probability memberships by comparing the distributions of $(\br,G,\varpi,\ \alpha,\delta,\vec{\mu})$ in two patches of the sky: one containing the LMC, one in a direction symmetric to the LMC, but flipped about the Milky Way midplane. We then model the LMC patch distribution as a sum of a \emph{pure} LMC distribution and a \emph{foreground} (FG), where the foreground model, $p_{FG}(\br,G,\varpi,\ \alpha,\delta,\vec{\mu}) \equiv  p_{sym}(\br)\cdot p_{sym}(G)\cdot p_{sym}(\varpi)\cdot p_{flat}(\alpha,\delta,\vec{\mu})$ is based on the colour, magnitude and parallax distribution in the symmetric patch ($p_{sym}$), and is uniform in sky position and proper motion ($p_{flat}$). From these two distributions, we can then work out the membership probabilities for the LMC. We retain stars with $\gtrsim 50$\% membership probability. We have tested that these membership probabilities yield results consistent with \cite{jimenez-arranz_2023}. The overall \MH\ selection function, the \MH\ quality cuts and the LMC membership leave a sample of \nsample\, stars.

\subsection{Stellar Ages for the Sample}

For part of the analysis below we also want to have observational constraints on the stellar ages, $\tau_*$. For stars with spectroscopic parameters, photometry, and good distances this can be done by isochrone fitting. For the LMC this has been shown by \citet{povick_2023a_ages}, drawing on SDSS-IV data. We have carried out an analysis for LMC stars with parameters from XP spectra (Povick et al, \emph{in prep.}) that conceptually follows this analysis quite closely, with only some implementation differences that speed up the modeling. We only summarize the salient features of the stellar age modeling. Any such modeling must be based on a set of isochrones, where we have adopted the PARSEC isochrones here \citep{PARSEC, marigo_2017_parsec}.
These predict a set of photometric (absolute) magnitudes and spectroscopic parameters.
For each star, one then finds the best \emph{parameters}, i.e. the age $\tau_*$, metallicity \MH\ and (initial) stellar mass, M$_{ini}$ 
that best predict the \emph{data}, i.e. the photometry (here the absolute magnitudes in $B,G,R,J,H,K$) and the spectroscopic parameters T$_\mathrm{eff}$ and \MH$_{XP}$. The best parameters are identified by minimizing the chi squared $\chi^2-2\log (p_{iso})$ across a fine set of pre-computed isochrone points, where $p_{iso}$ denotes the probability of any $(\tau_*,\MH,M_{ini})$ isochrone point which arises from the initial mass function $p(M_{ini})$ and the rate at which the observables $\overrightarrow{obs}$ change with $M_{ini}$, $\mathrm{d}{\overrightarrow{obs}}/\mathrm{d} M_{ini}$. The tests in Povick et al, (\emph{in prep.}) imply a typical age precision of $\Delta$log$\tau_*\sim 0.15$ dex.

%
%
%
%
%
%
\begin{figure*}
\includegraphics[width=\textwidth]{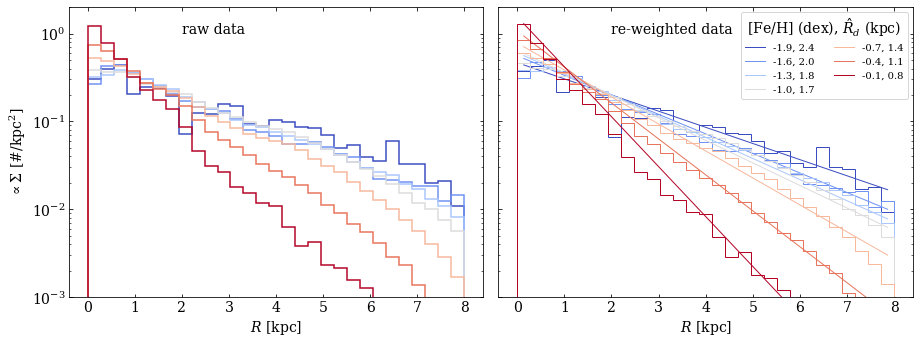}
\caption{Normalized surface density profiles of mono-[M/H] populations. Left: profiles constructed by basic source count of the data. Right: profiles constructed by the source count reweighted with the selection function's inverse (histogram), and best-fit exponential profiles found by maximizing Eq. \ref{eq:loglik} (lines). The lines are color-coded according to their metallicity, from -0.1 dex (red) to -1.9 dex (blue).}
\label{fig:surface_density}
\end{figure*}

\section{Maps of Mono-Abundance Populations}
\label{sec:global-MH-inspection}

We start by illustrating
metallicity maps of the LMC in Fig.~\ref{LMC-mono-abundance-maps}.
First, Fig.~\ref{LMC-mono-abundance-maps}a shows the average metallicity as a function of position on the sky. The central bar clearly stands out as being more metal-rich on average than the rest of the LMC. In the LMC disk, the metallicity continuously drops outwards.
The other panels provide a more detailed picture of the spatial distribution of mono-[M/H] populations: Fig.~\ref{LMC-mono-abundance-maps}b and~c show that there is no bar for $\MH<-1$, while there already is evidence for a disk at $-2<\MH<-1.5$. Instead, Fig.~\ref{LMC-mono-abundance-maps}d shows first evidence of the bar only at $\MH>-1$. We also note that weak evidence for spiral arms is visible already at $-1.5<\MH<-1$ (Fig.~\ref{LMC-mono-abundance-maps}c) before the bar appears, but fades away at $-0.5<\MH<0$ (Fig.~\ref{LMC-mono-abundance-maps}e) while the central bar persists. 
Finally, we note that Fig.~\ref{LMC-mono-abundance-maps}b-e have been corrected for completeness by dividing the raw number counts of stars by the completeness shown in Fig.~\ref{fig:LMC-completeness}.

%
%
%
%
%
%

\section{Spatial Structure Modeling of Mono-[M/H] populations}
\label{sec:spatial-structure-modelling}

\begin{figure}
    \centering
    \includegraphics[width=\columnwidth]{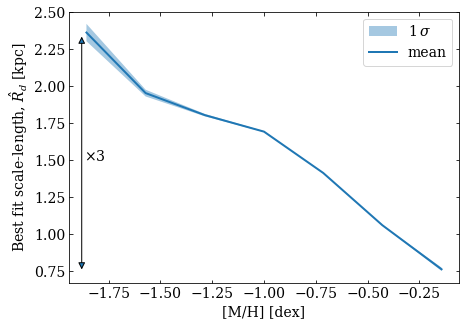}
    \caption{Best fit scale-length as a function of metallicity. The blue shade represents the 1-sigma region around the mean found from bootstrapping the data, and the solid line represents the mean scale-length. The scale-length changes by a factor 3 between the metal-poor and the metal-rich populations.}
    \label{fig:scale-lengths}
\end{figure}

We now turn to modelling the radial distribution of mono-[M/H] populations. We model the surface density profile as an exponential profile with a scale-length $R_d$, and quantify the scale-length of each mono-[M/H] population.

Figure \ref{fig:surface_density} shows the normalized surface density profiles of mono [M/H] populations. The profiles look roughly exponential, although the metal-rich profiles seem overdense in the central region (bar). The metal-poor populations are spatially extended, whereas the metal-rich populations are confined to the inner galaxy. To extract the scale-lengths of these populations, we model the surface density profile as
\begin{equation}
    \Sigma(R) = \Sigma_0 \exp(-R/R_d).
\end{equation}
In the rest of this work, we focus on extracting the scale-length $R_d$, and treat the normalization $\Sigma_0$ (of mono-abundance populations and later mono-age mono-abundance populations) as a nuisance parameter because fitting it would require to get the normalization of the underlying stellar populations and require isochrone modelling, which is outside the scope of this work.
The spatial distribution of the stars in the dataset does not represent that of the underlying population because of selection effects. Namely, in crowded regions, fewer stars made it to the value-added catalog. Therefore, the stars in the sample have a radial distribution that depends both on the underlying density and on a selection probability $S(\alpha, \delta)$,
\begin{equation}\label{eq:model_indivd}
    p(x,y) = C^{-1}(R_d)\frac{1}{R_d^2} \exp(-R/R_d)S(x,y),
\end{equation}
with the normalization constant
\begin{equation}
    C(R_d) = \iint \frac{1}{R_d^2} \exp(-R/R_d)S(x,y) dxdy
    \label{eq:volume}
\end{equation}
where in practice, we have evaluated $S$ as a function of sky position $S(\alpha(x,y),\delta(x,y))$ and the galactocentric radius is a function of Cartesian coordinates, $R(x,y) = \sqrt{x^2 + y^2}$.
Assuming the measurements of the positions are independent, we can build a likelihood function for $R_d$ as the joint distribution of the dataset (i.e. the product of individual probabilities of Eq. \ref{eq:model_indivd}) given the model parameters. The log-likelihood is 
\begin{equation}\label{eq:loglik0}
\begin{split}
    \ln \mathcal{L}(\{x_i,y_i\};R_d) = \sum_{i=1}^{N_\star} \biggl[ &-\ln(R_d^2) -\frac{R_i}{R_d} \\
    &+ \ln(S(x_i,y_i)) - \ln(C) \biggr].
    \end{split}
\end{equation}
We aim to search for the value $\hat{R}_d$ of the scale-length parameter $R_d$ that maximizes the joint probability of the dataset under that model $\hat{R}_d = \underset{R_d}{\argmax} \ln \mathcal{L}(\{x_i, y_i\}; R_d)$, i.e. the $R_d$ that maximizes Eq. \ref{eq:loglik0}.
The log-likelihood maximizes with respect to $R_d$ if the following quantity is maximized (dropping constants)
\begin{equation}
    \ln \mathcal{\tilde{L}}(\{x_i,y_i\};R_d) =-2\ln(R_d)  -  \ln(C) -\frac{\langle R \rangle}{R_d},
    \label{eq:loglik}
\end{equation}
where $\langle R \rangle = \frac{1}{N}\sum_i R_i$, and we remind the reader that $C = C(R_d)$, i.e. it is not a constant to drop out of the likelihood.
The selection function, $S(\alpha, \delta)$, is derived in Appendix~\ref{appendix:completeness} and shown in Fig.~\ref{fig:LMC-completeness}.

\begin{figure*}
    \centering
    \includegraphics[width=\textwidth]{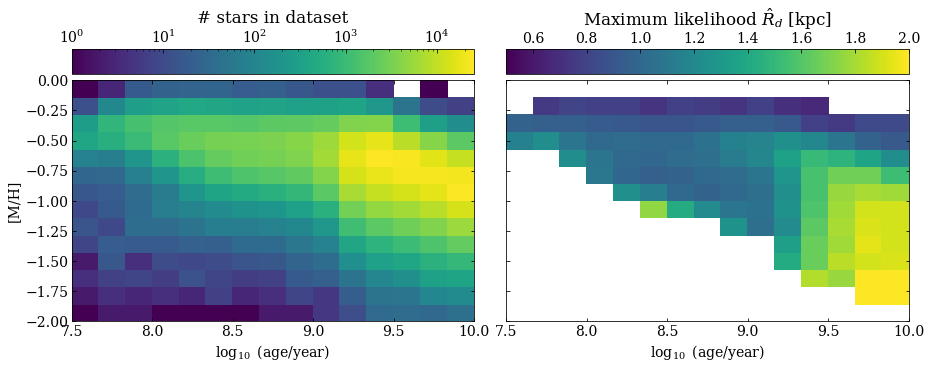}
    \caption{\label{age-metallicity-structure}From left to right: number of stars in the data set in age-metallicity bins and best fit LMC scale-length for mono-age mono-metallicity populations, $\hat{R}_d(\tau_\star, \MH)$. This plot shows the age- and metallicity- gradients in the LMC, reflecting an outside-in formation scenario with the younger and metal-richer stars more centrally concentrated. Young stars do not seem to exhibit any strong metallicity gradient, whereas older stars do. We only display best-fit scale-lengths for age-metallicity bins having 100 stars or more.}
    \label{fig:enter-label}
\end{figure*}

In principle, the selection function depends on color too, but we decide to simplify it by modeling it only in the 2D spatial position. The two main reasons for this choice are that, first, some stars (about 0.6\%) have no BP or RP photometry in the Gaia catalog, such that we cannot infer completeness via ratio-ing (see appendix) and, second, the \texttt{gaiaunlimited} package \citep{2023A&A...669A..55C} at the time of analysis only supports a dependence on apparent $G$ magnitude but not on any colour. Therefore, we have no practical means for introducing a colour dependence. 

The previous eyeballing Fig. \ref{fig:surface_density} helps us to know that $R_d$ should be in the range of [$0.5,3.0$] kpc. So, we pre-calculate Eq.~\ref{eq:volume} on a grid of $R_d$ values in that range and interpolate to fit the likelihood function (this is not strictly necessary, but it accelerates the calculations that follow).
Since the likelihood function is only one-dimensional, finding the maximum can be done directly. We brute-force evaluate the log-likelihood on a grid of $R_d$ and find the maximum value on the grid. We do this in metallicity bins and bootstrap the samples 100 times to obtain uncertainties in the best-fit estimates, presented in Fig.~\ref{fig:scale-lengths}. When taking the full populations to estimate a global scale-length, we find $R_d = 1.5$ kpc.

%
%
%
%
%
%

\section{Spatial Structure Modeling of mono-age mono-[M/H] populations}
\label{sec:spatial-structure-modelling-age-mh}

We repeat the analysis above, with an additional dimension provided by the stellar ages derived from isochrone fitting by Povick et al. (\emph{in prep}), in order to address the question: \emph{at present, where are stars that presumably formed at the same time, and from similarly-enriched gas (so most likely at similar locations)?}

\begin{figure*}
    \centering
    \includegraphics[width=\textwidth]{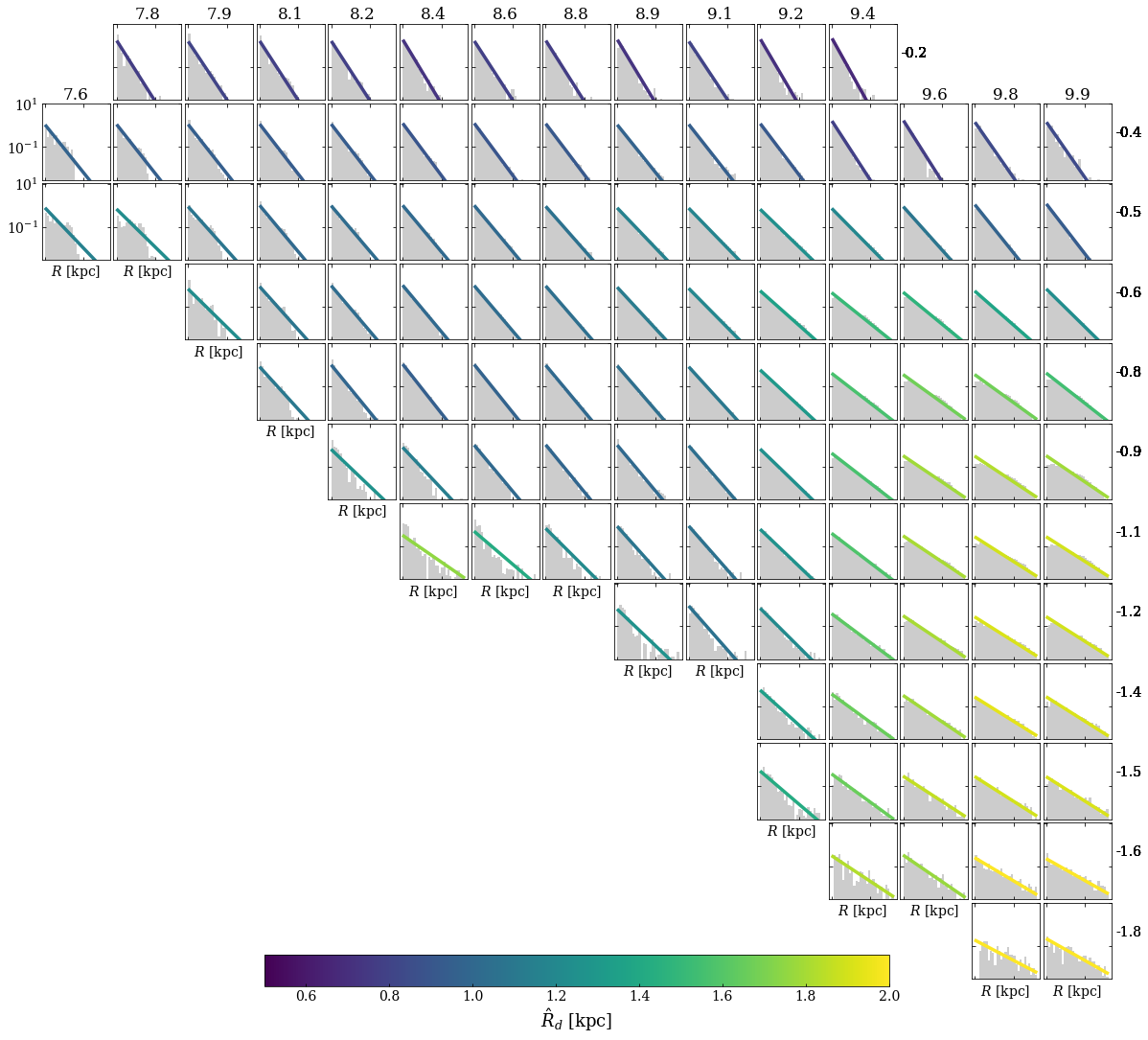}
    \caption{Normalized radial distribution of the sample in the age-metallicity bins represented by each panel. Age increases to the right and metallicity increases upwards. The distributions are re-weighted by the selection fractions $1/S_i$ and the Jacobian $1/R$ to represent the surface density profile (grey histograms), and best-fit exponential profiles (colored solid lines) as presented in the middle panel of Fig. \ref{age-metallicity-structure}. The color code is by best-fit scale-length $\hat{R}_d$ (0.5 kpc in blue to 2 kpc in yellow). The values of the ages are annotated as titles above each column, and metallicities are annotated on the right side of each line. The profiles are remarkably exponential in most age-metallicity bins.
    \label{fig:radial_structure_age_metallicity}}
\end{figure*}

We assume that the spatial selection function is valid in age and metallicity bins, i.e.\ the sample's completeness is the same function of color and magnitude (i.e.\ indirectly age and metallicity) as a function of position, so that all stellar populations have the same spatial dependence, up to a normalization factor. Then, we can maximize Eq.~\ref{eq:loglik} as before. The scale-lengths are depicted in Fig.~\ref{age-metallicity-structure}, and the near-exponential radial profiles of mono-age mono-\MH\ populations are shown in Fig.~\ref{fig:radial_structure_age_metallicity} along their best-fit exponential profiles.

We find that for ages of 1 Gyr and greater, the scale-length decreases with metallicity from 2 kpc at \MH$\sim$-2 dex to 1 kpc at \MH$\sim$-0.3 dex. For younger populations, the scale-length is independent of metallicity.

%
%
%
%
%
%
\section{Discussion and Conclusion \label{section:discussion}}

Using the XGBoost models from \citet{Andrae_rix_chandra_2023}, we have derived empirical \MH\ estimates and isochrone ages for \nsample\, giant stars that are likely members of the LMC.
Based on these metallicities and then the ages, we investigate the radial density structure of the LMC in mono-abundance (\MH ) and mono-age (log$\tau_*$) stellar populations.

We find that the radial structure of the LMC is simple: the radial mean surface density profiles for basically all mono-abundance stellar populations is well described by a single exponential. The scale-lengths vary strongly, but smoothly and systematically with \MH , in the sense that the most metal-poor populations are most extended, and the most metal-rich populations are far more centrally concentrated (see Fig.~\ref{fig:surface_density} and \ref{fig:radial_structure_age_metallicity}), with a three times shorter scale-length (see Fig.~\ref{fig:scale-lengths}). 

When considering stellar populations that are both mono-age and mono-metallicity, the LMC's azimuthally averaged radial profile still close to exponential. At given age, the exponential scale-lengths decrease with increase metallicity, but only at  ages $>$ 1 Gyr; the constant with metallicity in younger populations. 

The age- and abundance-dependent radial structure of the stars, $R_d(\MH ,\mathrm{log}\tau)$ has not been determined before. However, the implied metallicity and age gradients are consistent with prior literature, that finds that young stellar populations, young star clusters \citep[$<$ 1 Gyr][]{palma_2015}, and the HII regions in the gas present no noticeable metallicity or abundance gradient \citep{kontizas_1993, dufour_1975} whereas studies targeting older stellar populations or clusters do present metallicity gradients \citep{kontizas_1993,cioni_2009_gradients, povick_2023b_abundance_gradients}.

The stellar metallicity gradient in the older populations at face value implies that there probably was an analogous gradient in the stars' birth gas.  However, if the gas was azimuthally well mixed, and if there was a gas metallicity gradient, then we should expect that a star's age and metallicity is informative about its birth radius as found in the Milky Way \citep[e.g.,][]{minchev_2018, frankel_2018, frankel_etal_2019, frankel_2020} and partly in simulations of LMC-mass galaxies \citep{lu_2024_LMCbirthradii}. 

However, the present-day radial density distributions of mono-age mono-metallicity populations do not peak around specific radii, as seen in the Milky Way \citep[e.g.][]{bovy_etal_2016}. Their extended, near-exponential profiles would suggest a strong radial redistribution of the stars between their births and present. For example, their orbits could be eccentric in a random or ordered way (increased dispersion vs tidal perturbation) or the stars could have migrated radially. Other possibilities could be that the ISM of the LMC was not well-mixed at the formation of the stars, so that stars of various metallicities could be born at the same radius and time. 

The absence of radial metallicity gradients in the young populations seems to imply that if there used to be a metallicity gradient in the gas in the past, but if there is none today, the old gas might have been replaced by new one uniformly accreted. This could happen e.g., during a collision between the LMC and the SMC, although current constraints from \cite{choi_2022, zivick_2018} seem to place a more recent ($\leq 250$ Myr) collision between the two galaxies, so this could probably not fully explain the absence of metallicity gradient starting for the 1 Gyr old population.

The increasing presence of younger stars at smaller radii could mean either \emph{outside-in} growth, or \emph{outside-in interruption of star formation}, as also suggested by \cite{meschin_2014, gallart_2008} when finding that younger stars are more centrally concentrated. The presence of metallicity gradients in stars $>1$ Gyr, and its absence in younger stars,  may be evidence for the recent perturbing event between the SMC and the LMC. For example, it would seem plausible that along its way to its pericenter with the Milky Way, the LMC got stripped of its gas in an  outside-in regime as reflected by its age gradient, as dwarf galaxies can do when approaching their host and undergo outside-in ram pressure stripping \citep[e.g.,][]{rohr_2023, schaefer_2017}. Subsequent fresh gas accretion (possibly by the SMC, although the timing would need to be clarified) could possibly have provided fresh gas, deprived of a radial metallicity gradient, from which the younger population of stars formed \citep[as suggested in e.g.,][]{rubele_2012}.
The scenario described above seems a-priori plausible. However, it does not explain why mono-age mono-\MH\ populations are still spatially extended with a near-exponential radial profile. Others have pointed to a similar puzzle, finding that at given age and position in the LMC, the there was a wide spread of metallicities. What drove this? Orbit evolution, an ISM that was not well mixed, or one that had azimuthal variations?

While we have examined the azimuthally averaged radial profile of the LMC here,  it is clear that the LMC is not axisymmetric: it lopsided with a prominent star-forming spiral arm and has a central bar. 
A useful next step will be to forward model the possible physical mechanisms setting the star formation history, chemical enrichment, and radial structure of the LMC and the SMC. The model could for example predict the radially-dependent star formation history of the LMC and SMC as they fall into the Milky Way halo (and possibly get partially ram pressure stripped), and their possible interaction. Variations of this model could be constrained against the dataset. This is beyond the scope of the current work but a promising future avenue towards more quantitative answers.

%
%
%
%
%
%
\section{Acknowledgements}

This work has made use of data from the European Space Agency (ESA) mission {\it Gaia} (\url{https://www.cosmos.esa.int/gaia}), processed by the {\it Gaia} Data Processing and Analysis Consortium (DPAC, \url{https://www.cosmos.esa.int/web/gaia/dpac/consortium}). Funding for the DPAC has been provided by national institutions, in particular the institutions participating in the {\it Gaia} Multilateral Agreement.
NF was supported by the Natural Sciences and Engineering Research Council of Canada (NSERC), funding reference number CITA 490888-16, through a CITA postdoctoral fellowship, and acknowledges partial support from an Arts \& Sciences Postdoctoral Fellowship at the University of Toronto.
NF thanks David Nidever for interesting discussions during the Flight C of the \href{https://galaxy-formation-meeting.org/}{Wide Field Spectroscopy vs Galaxy Formation Theory meeting} held at the Biosphere 2, Arizona.
NF thanks Gurtina Besla for an interesting discussion at the University of Arizona at early stages of this work.
NF thanks Juna Kollmeier, Scott Tremaine, David W. Hogg, Arnaud Siebert and Benoit Famaey for stimulating discussions on this project, and Jo Bovy and his group for an interesting group meeting discussion.
\bibliography{lit}

\appendix

\section{Validation of \MH$_{XP}$\ against Extant LMC Spectroscopy}\label{appendix:XPvalidation}

\begin{figure}
\includegraphics[width=0.47\textwidth]{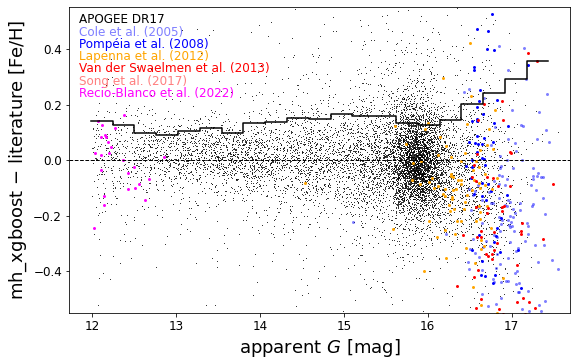}
\caption{Validation of empirical \MH\ estimates from \citet{Andrae_rix_chandra_2023}. Black dots represent cross-validated results for 9\,443 stars from APOGEE~DR17 that are LMC members. The other results are independent validation. The black step function indicates the standard deviation of all residuals binned by apparent $G$. As we approach the faint-end publication limit $G=17.65$ of XP spectra in Gaia~DR3, the \MH\ estimates become increasingly noisy but show no bias.}
\label{fig:LMC-MH-validation-vs-apparent-G}
\end{figure}

\citet{Andrae_rix_chandra_2023} carefully validated their \MH$_{XP}$\ estimates against large spectroscopic surveys, star clusters, wide binaries and solar analogues across the entire sky. However, their validation efforts mostly focused on the bright end, whereas in the present work on the LMC we are very much at the faint-end limit of the data. Therefore, Fig.~\ref{fig:LMC-MH-validation-vs-apparent-G} presents a brief additional validation specific to the LMC. First, we compare our [M/H] values to the subset of APOGEE~DR17 stars that reside in the LMC. Second, we also compare our results to independent spectroscopic metallicity estimates \citep{2005AJ....129.1465C,2008A&A...480..379P,2012ApJ...761...33L,2013A&A...560A..44V,2017AJ....153..261S}. These probe to fainter stars than the APOGEE~DR17 training sample. Third, we compare our metallicities to  spectroscopic metallicity estimates from \citep{2022arXiv220605541R} derived from a Gaia RVS spectra. These are few and focus on the bright end, given the RVS limit in Gaia~DR3.

While the empirical \MH\ estimates become increasingly noisy towards the faint end, they remain unbiased. Assuming that the ground-based spectroscopic measurements have negligible errors compared to the empirical \MH\ estimates, the 141 stars with $G>17$ have a standard deviation of 0.311, which is slightly better than what Fig.~9 in \citet{Andrae_rix_chandra_2023} suggests due to replacing the noisy parallax by the LMC distance. This clearly demonstrates that the empirical \MH\ estimates remain consistent in the LMC. Consequently, the increased random noise due to the LMC stars being relatively faint can be compensated through averaging over large numbers of stars without accumulating systematic errors. However, since it is plausible to assume that there is a steep gradient in the intrinsic \MH\ distribution of the LMC, large noise could lead to an asymmetric scattering of high-\MH\ stars contaminating low-\MH\ samples. Therefore, motivated by Fig.~\ref{fig:LMC-MH-validation-vs-apparent-G}, we add another quality cut of $G<17.65$ in order to reduce the impact of noisy \MH\ estimates. This reduces our total LMC sample size from $\sim$1.6~million down to $\sim$0.7~million.

\section{Completeness considerations}
\label{appendix:completeness}

We seek the completeness of \MH\ estimates from XGBoost, which is $P({\rm has\,\MH})$. First, we introduce the binary variable of being in Gaia~DR3 or not via marginalisation:
\begin{equation}
P({\rm has\,\MH})
=
P({\rm has\,\MH} \;,\; {\rm in\,Gaia})
\end{equation}
\begin{equation}
\qquad
+ \underbrace{P({\rm has\,\MH} \;,\; {\rm not\,in\,Gaia})}_{=0}
\end{equation}
If a source is not in Gaia, it cannot have \MH\ from XGBoost, such that the second term drops out. Next, we introduce the binary variable of having XP spectra in Gaia~DR3 or not via marginalisation:
\begin{equation}
P({\rm has\,\MH})
=
P({\rm has\,\MH} \;,\; {\rm in\,Gaia}) 
\end{equation}
\begin{equation}
\qquad
=
P({\rm has\,\MH} \;,\; {\rm has\,XP}\;,\; {\rm in\,Gaia}) 
\end{equation}
\begin{equation}
\qquad
+ \underbrace{P({\rm has\,\MH} \;,\; {\rm has\,no\,XP}\;,\; {\rm in\,Gaia})}_{=0}
\end{equation}
Again, if a source is in Gaia~DR3 but has no XP spectrum, it cannot have \MH\ results from XGBoost, such that the second term drops out, too. Finally, we introduce the condition of having CatWISE $W_{1/2}$ photometry, which is required as input feature for XGBoost to produce \MH\ estimates:
\begin{equation}
P({\rm has\,\MH})
=
P({\rm has\,\MH} \,,\; {\rm has\,XP}\,,\; {\rm in\,Gaia})
\end{equation}
\begin{equation}
\qquad
=
P({\rm has\,\MH} \,,\; {\rm has\,W_{1/2}} \,,\; {\rm has\,XP}\,,\; {\rm in\,Gaia}) 
\end{equation}
\begin{equation}
\qquad
+ \underbrace{P({\rm has\,\MH} \,,\; {\rm has\;no\,W_{1/2}} \,,\; {\rm has\,XP}\,,\; {\rm in\,Gaia}) }_{=0}
\end{equation}
Let us rewrite this using conditional probabilities:
\begin{equation}
P({\rm has\,\MH})
=
P({\rm has\,\MH} \;|\; {\rm has\,W_{1/2}}\,,\;{\rm has\,XP}\,,\; {\rm in\,Gaia}) 
\end{equation}
\begin{equation}
\qquad
\times\, P({\rm has\,W_{1/2}} \;|\; {\rm has\,XP}\,,\; {\rm in\,Gaia}) 
\end{equation}
\begin{equation}
\qquad
\times\, P({\rm has\,XP}\;|\; {\rm in\,Gaia}) \,\times\, P({\rm in\,Gaia})
\end{equation}
Obviously, if a source has XP spectra, it also must be in Gaia~DR3, such that ``in Gaia'' can be omitted from the first factor.
Given the increased noise in \MH\ estimates (c.f.~Fig.~\ref{fig:LMC-MH-validation-vs-apparent-G}), we limit everything to $G<17.65$ in order to eliminate faint XP spectra published in Gaia~DR3 for QSOs, galaxies and UCDs. Thus, $G<17.65$ becomes a ``spectator variable'':
\begin{equation}\label{eq:completeness-factorised}
\begin{split}
P({\rm has\,[M/H]}\;|\; G<17.65) \\
\qquad
=
P({\rm has\,[M/H]} \;|\; {\rm has\,W_{1/2}}\,,\; {\rm has\,XP}\;,\;G<17.65) \\
\times P({\rm has\,W_{1/2}} \;|\; {\rm has\,XP}\,,\; G<17.65)  \\
\qquad\phantom{=}
\times P({\rm has\,XP}\;|\; {\rm in\,Gaia}\;,\;G<17.65)  \\
\qquad\phantom{=}
\times P({\rm in\,Gaia}\;|\;G<17.65)
\end{split}
\end{equation}
Specifically, $P({\rm has\,\MH} \;|\; {\rm has\,W_{1/2}}\,,\; {\rm has\,XP}\;,\;G<17.65)$ may be less than 1 because a source may have no parallax, may have no $G_{\rm BP}$ or $G_{\rm RP}$ despite having XP spectra,\footnote{In fact, there are 4 sources in Gaia~DR3 that have XP spectra, $G<17.65$ and a parallax but which do not have $G_{\rm BP}$. There is no such case for $G_{\rm RP}$.} or its XP spectrum may have a low signal-to-noise ratio such that one or more of its synthesised fluxes become negative.

\begin{figure*}
\includegraphics[width=0.47\textwidth]{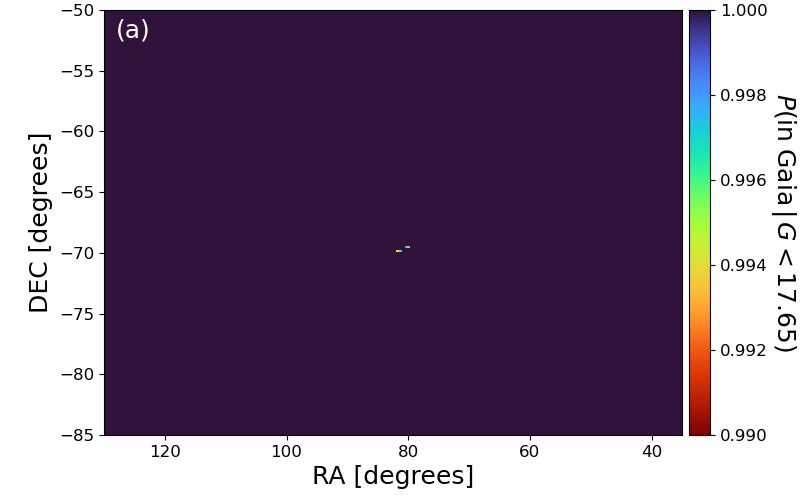}
\includegraphics[width=0.47\textwidth]{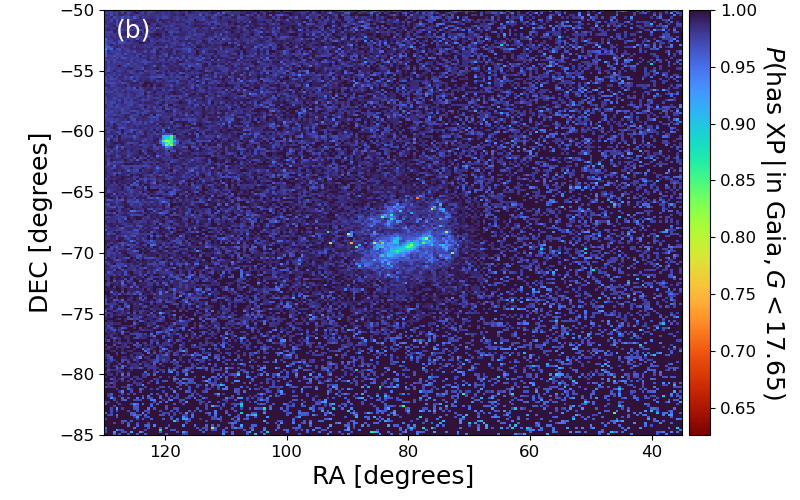}\\*
\includegraphics[width=0.47\textwidth]{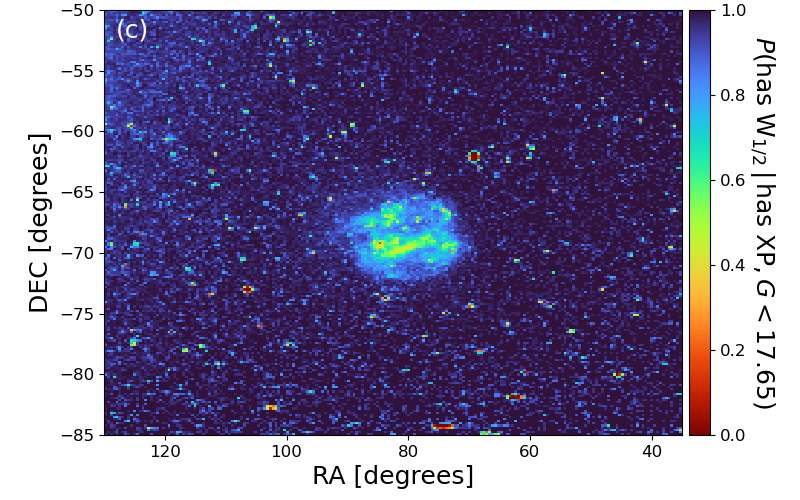}
\includegraphics[width=0.47\textwidth]{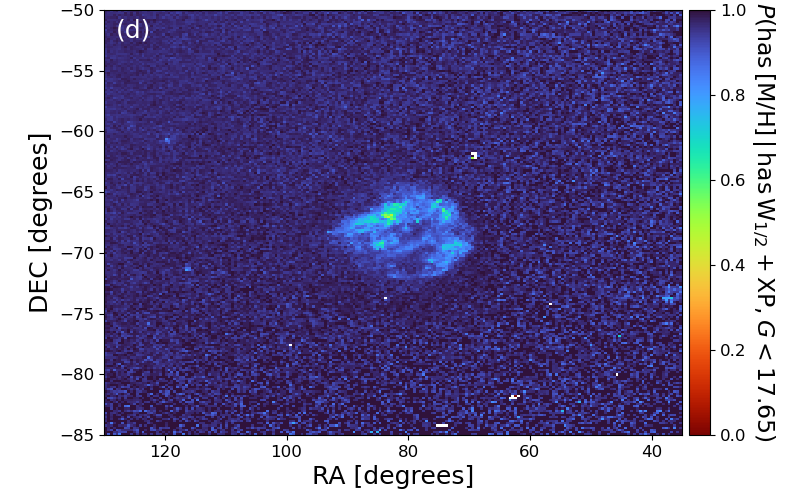}
\caption{Completeness of the LMC according to the four factors in Eq.~\ref{eq:completeness-factorised}. Panel (a) represents the last term, i.e. the probability that a star with $G < 17.5$ is in Gaia; panel (b) represents the second to last term, i.e. the probability to have an XP spectra conditioned on (a); panel (c) the second term, i.e. whether Catwise has $W_1$ and $W_2$ entries conditoinned on (a-b); and panel (d) the first term, i.e. whether a metallicity could be estimated conditioned on (a-b-c). Note the drastically different colour-bar ranges. The resulting selection function is the multiplication of those four panels and is displayed in Fig.~\ref{fig:LMC-completeness}.}
\label{fig:completeness-step-by-step}
\end{figure*}

The first three factors in Eq.~(\ref{eq:completeness-factorised}) can be easily obtained from taking ratios of number counts in the Gaia catalog. As is evident from Fig.~\ref{fig:completeness-step-by-step}c, there is a substantial loss of completeness of stars actually having XP spectra in the bar region of the LMC, which is due to missing CatWISE photometry. As Fig.~\ref{fig:completeness-step-by-step}d shows, the loss of completeness due to missing parallax or noisy XP spectra leading to negative synthesised fluxes is small.  Conversely, Fig.~\ref{fig:completeness-step-by-step}b shows that most sources with $G<17.655$ have an XP spectrum, though there also is a minor loss of completeness in the innermost part of the LMC.
The fourth factor in Eq.~(\ref{eq:completeness-factorised}), $P({\rm in\,Gaia}\;|\;G<17.65)$, can be obtained via \texttt{gaiaunlimited} \citep{2023A&A...669A..55C}. Fig.~\ref{fig:completeness-step-by-step}a shows that according to \texttt{gaiaunlimited} the completeness of Gaia at $G<17.65$ is perfect everywhere. Consequently, it is acceptable to consider Gaia as being fully complete at $G<17.65$ within the LMC footprint and only consider the first three factors in Eq.~(\ref{eq:completeness-factorised}), whereof the second factor (CatWISE completeness) clearly dominates over the first and third factor.

\end{document}